\documentstyle[11pt,newpasp_charbonneau,twoside,epsf]{article}
\def\edcomment#1{\iffalse\marginpar{\raggedright\sl#1\/}\else\relax\fi}
\reversemarginpar

\begin{document}
\title{Present and Near-Future Reflected Light Searches for Close-In Planets}
\author{David Charbonneau$^{1,2}$ and Robert W. Noyes$^{1}$}
\affil{$^1$Harvard-Smithsonian Center for Astrophysics, 60 Garden St., Cambridge, MA 02138, USA}
\affil{$^2$High Altitude Observatory, National Center for Atmospheric 
Research, P. O. Box 3000, Boulder, C0 80307, USA}

\begin{abstract}
Close-in extrasolar giant planets may be directly detectable
by their reflected light, due to the proximity of the planet
to the illuminating star. 
The spectrum of the system will contain
a reflected light component that varies in amplitude and Doppler shift
as the planet orbits the star.  Intensive searches for this effect
have been carried out for only one extrasolar planet 
system, $\tau$~Bo\"o.
There exist several other attractive targets, including the transiting
planet system HD~209458.
\end{abstract}

\section{Introduction and Motivation}
Radial velocity surveys of nearby Sun-like stars have 
uncovered a population of close-in orbiting companions
of roughly Jupiter mass.  The ten such objects with
semi-major axes \mbox{$a \la 0.1$ AU} are listed in Table~1, along 
with the values for the semi-major axes, (minimum) masses, 
spectral types of the stars, and equilibrium temperatures 
(calculated from the estimated values 
for the semi-major axes, stellar radii and effective temperatures, and
assuming a Bond albedo of $A$).    

\begin{table}
\caption{Close-In Planets Ordered by Increasing Semi-Major Axis}
\begin{tabular}{l c c c c}
\tableline
Star & Spectral Type & $a$ (AU) & $M_p \ (M_{\rm Jup})$ & $T_{\rm eq} (1-A)^{-\frac{1}{4}}$ (K) \\ 
\tableline
HD 187123 & G3~V & 0.042 & 0.52/$\sin i$ & 1400 \\
HD 75289 & G0~V & 0.046 & 0.42/$\sin i$ & 1600 \\
$\tau$ Bo\"o & F7~V & 0.046 & 3.87/$\sin i$ & 1600 \\
HD 209458 & G0~V & 0.047 & 0.69 & 1500 \\
51 Peg & G2~V & 0.051 & 0.45/$\sin i$ & 1300 \\
$\upsilon$ And & F8~V & 0.059 & 0.71/$\sin i$ & 1500 \\
HD 217107 & G7~V & 0.072 & 1.28/$\sin i$ & 1000 \\
HD 130322 & K0~V & 0.088 & 1.08/$\sin i$ & 800 \\
55 Cnc & G8~V & 0.11 & 0.84/$\sin i$ & 700 \\
Gl 86 & K0~V & 0.11 & 3.6/$\sin i$ & 700 \\
\tableline
\tableline
\end{tabular}
\end{table}

A successful spectroscopic detection of an extrasolar planet in reflected
light would
yield the inclination, and hence the planetary mass, and would
also measure a combination of the planetary radius and albedo.
Furthermore, it would open the way to direct investigation of the 
spectrum of the planet itself.  Conversely, a low enough upper
limit would provide useful constraints on the radius and
albedo of the companion.

The predicted albedo of a close-in extrasolar giant planet
has been the focus of recent theoretical work 
(Marley et al. 1999; Seager, Whitney, \& Sasselov 2000),
and the possible values range by several orders of magnitude
based on the atmospheric temperature, chemical composition 
(see, e.g., Burrows \& Sharp 1999),
and the presence (or absence) of atmospheric condensates, and
their respective size distributions.

\section{The Method of Reflected Light}
The amplitude, relative to the star, 
of the observed flux reflected from a close-in
planet viewed at a phase angle $\alpha$ (where $\alpha$ is the angle
between the star and the observer as viewed from the planet) is
given by 
\begin{equation}
f_{\lambda}(\alpha) = \epsilon \ \phi_{\lambda}(\alpha) = p_{\lambda} \, \left( \frac{R_{p}}{a} \right)^2 \phi_{\lambda}(\alpha),
\end{equation}
where $R_{p}$ is the planetary radius, $a$ is the semi-major axis, 
and $p_{\lambda}$ is the wavelength-dependent
geometric albedo.  The quantity $\epsilon$ is the contrast ratio
at opposition.  The phase function $\phi_{\lambda}(\alpha)$ is the 
brightness of the planet viewed at an angle $\alpha$ relative to 
its value at opposition (where
by definition \mbox{$\phi_{\lambda}(0) = 1$}), 
and is a monotonically decreasing function
of $\alpha$ for all physically reasonable atmospheres.  
For planets detected by the radial velocity technique alone, 
$a$ is determined, and $\alpha$ at a given time is prescribed
by the known orbital phase $\Phi$ and the unknown orbital inclination $i$, 
but $R_{p}$, $p_{\lambda}$ and $\phi_{\lambda}(\alpha)$ remain unknown.

A distant observer viewing the solar system in the $V$ band would measure
a tiny reflected light ratio of \mbox{$f \simeq 4 \times 10^{-9}$} from Jupiter
viewed at opposition.  This ratio increases 
dramatically for close-in, giant planets, due to the small orbital separation.
Scaling equation~(1), we find
\begin{equation}
f_{\lambda}(\alpha) \simeq 9.1 \times 10^{-5} \ p_{\lambda} \, \left(\frac{R_{p}}{R_{\rm Jup}}\right)^{2} \left(\frac{0.05 \rm{AU}}{a}\right)^{2} \phi_{\lambda}(\alpha).
\end{equation}
This photometric modulation is most likely beyond the reach of
ground-based observations, but will be accessible to 
upcoming photometric satellite 
missions, such as COROT (see, e.g., 
Rouan et al. (1999)), provided such missions can achieve
a precision of \mbox{$\sim 20 \ \mu{\rm mag}$} with stability over 
timescales of a few days.

Photometry alone will not yield the orbital inclination, and hence the
planetary mass will remain constrained only by the lower limit
imposed by the radial velocity observations.  However, a spectrum
of a close-in planet system will contain a
secondary component that varies in brightness according to equation~(1)
and in Doppler shift (relative to the star) according to
\begin{equation}
v_{p}(\Phi) = - \ K_{s} \ \frac{M_{s}+M_{p}}{M_{p}} \ \cos{2 \pi \Phi},
\end{equation}
where $M_{s}$ and $M_{p}$ are the stellar and planetary masses.
The stellar radial velocity amplitude $K_{s}$ and
orbital phase $\Phi$ are determined from the observed radial velocity orbit
of the star.  For the close-in planets, 
\mbox{$|v_{p}| \simeq 100 \ {\rm km \, s^{-1}}$}, which is two 
orders of magnitude larger than the resolution of current spectrographs.
If the reflected light spectrum of the planet is detected,
then the planetary mass is determined by equation~(3).

Furthermore, as first discussed in Charbonneau, Jha \& Noyes (1998), if the
star has been tidally spun-up so that its rotation period is equal
to the planetary orbital period, then the planet would reflect a
non-rotationally-broadened spectrum, further distinguishing these
features from the stellar lines.

In summary, the
observational challenge is similar to that of transforming a single-lined
binary system into a double-lined system, in the case of an exceptionally
large contrast ratio between the two components.  
See Charbonneau et al. (1999), hereafter C99, for more details.

\begin{figure}
\plotone{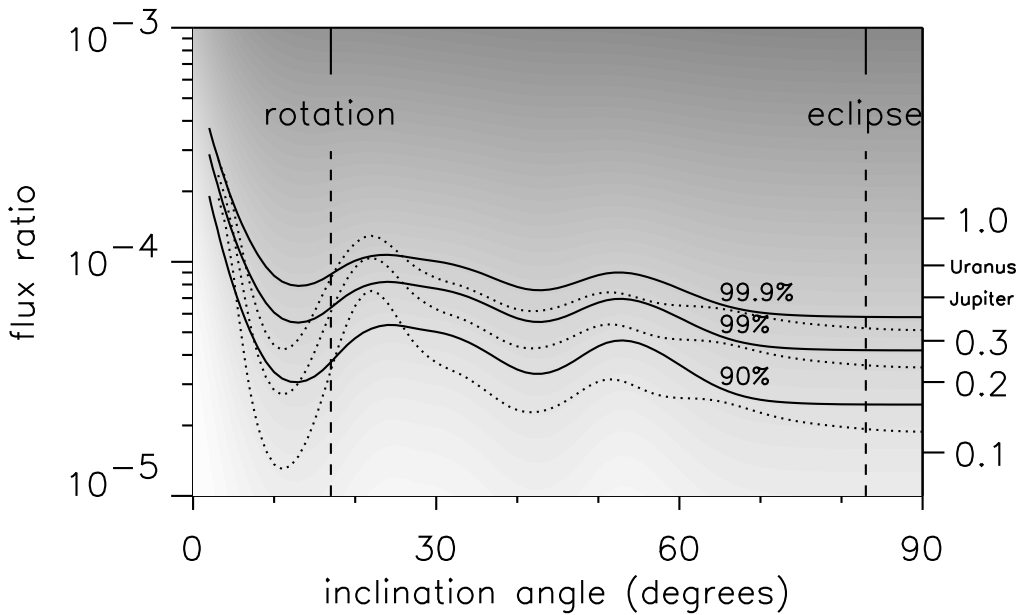}
\caption{The solid curves are the confidence levels
on the upper limit for the relative reflected flux $\epsilon$ 
in the $\tau$ Bo\"o system as
a function of orbital inclination $i$, if the 
reflected light is a copy of the stellar spectrum. 
The dashed curves are the same confidence levels under the
assumption that the system is tidally locked and thus
the planet reflects a non-rotationally-broadened copy of the 
stellar spectrum.  
Upper limits on the geometric albedo $p$ under the assumption that
\mbox{$R_{p} = 1.2 \ R_{\rm Jup}$} are shown on the right-hand axis,
and the values for Jupiter and Uranus are included for comparison. 
The lack of transits (Baliunas et al. 1997) excludes 
\mbox{$i \ga 83\deg$}, and \mbox{$i \la 17\deg$}
can be excluded under the assumption that the stellar rotation axis is 
co-aligned with that of orbital motion, as it would imply an abnormally
large rotational velocity (Gray 1982).}
\end{figure}

\section{\begin{boldmath}Results for $\tau$~Bo\"o\end{boldmath}}
Figure~1 summarizes our search for the reflected light spectrum
from the planet orbiting the star $\tau$~Bo\"o, as described in
C99.  At the 99\% confidence level, we
find no evidence for a reflected flux ratio in excess of 
\mbox{$1 \times 10^{-4}$}.
For edge-on values of the inclination \mbox{($i \ga 70\deg$)}, this ratio
is further restricted to be less than \mbox{$5 \times 10^{-5}$}.  
Assuming a planetary
radius of \mbox{$1.2 \ R_{\rm Jup}$} (Guillot et al. 1996), this limits the
geometric albedo to \mbox{$p_{\lambda} \le 0.3$} 
for \mbox{466 nm $\le \lambda \le$ 498 nm}.

These conclusions are marginally in conflict with the recent claim of
a possible detection of reflected light for the same system by
Cameron et al. (1999).  They find a reflected light ratio of
\mbox{$(1.9 \pm 0.4) \times 10^{-4}$} in the wavelength range from
456 nm to 524 nm, and do not detect the planet in wavelength bands
on either side of this region (385 nm to 456 nm, and 524 nm to 611 nm).
Figure~2 demonstrates the confidence 
intervals that we would have obtained on the two unknowns 
\mbox{$\{\epsilon,i\}$} for inclinations near the one claimed by
Cameron et al. (1999) and for a selection of contrast ratios.

Taking into account the different descriptions of the assumed phase
function $\phi_{\lambda}(\alpha)$, and the respective errors stated in the
two papers, it is possible that the two results are 
consistent, i.e. the inclination
claimed by Cameron et al. (1999) is correct, but the amplitude of 
the signal is less than their most probable value 
by 1.5 times their quoted error.  
However, Cameron et al. (1999)
caution that there is a 5\% possibility that their detection is spurious.
This controversy should be resolved by observations in the
near future that can now be tailored (in selection of wavelength 
region and orbital phase) to verify the claimed detection.

\section{Additional Targets}
There are two principal considerations that enter into the choice of which
planets may be most easily detected by reflected light:  An
ideal system is one with a small semi-major axis and a large stellar 
apparent brightness, and thus a large reflected light signal relative 
to the photon noise of the star.  There are two other systems 
that satisfy these criteria, and we compare below the prospects for
detecting each of these with those for $\tau$~Bo\"o (\mbox{$a = 0.046$ AU},
\mbox{$V = 4.5$}).

The planet orbiting $\upsilon$~And is further from its star 
\mbox{($a = 0.059$~AU)}, 
but the star has a greater apparent brightness \mbox{($V = 4.1$)}.  
Combining these two effects,
the detection threshold that could be established 
(given an equivalent amount of observing time) 
would be 1.36 times higher relative to the primary star
than that for $\tau$~Bo\"o.  For the planet orbiting 51~Peg, 
the larger semi-major axis \mbox{($a = 0.051$~AU)} and fainter star 
\mbox{($V = 5.5$)} result in
a detection threshold that is 1.93 times higher than that for $\tau$~Bo\"o. 

Near edge-on inclinations
\mbox{($i \simeq 90\deg$)} are desirable for reflected light
observations since they allow the planet to be viewed
at the full range of phase angles.
There are additional considerations that may constrain the inclination
for these two systems.
Assuming that the orbital planes
of the three planets orbiting
$\upsilon$ And (Butler et al. 1999) are co-aligned, stability arguments
(Laughlin \& Adams 1999) favor \mbox{$\sin{i} \ga 0.75$}.  
However, {\it Hipparcos} astrometry 
favors (Mazeh et al. 1999) \mbox{$\sin{i} \simeq 0.4$}.  The rotation
periods of $\upsilon$~And  and 51~Peg indicate that the stellar
rotation has not become tidally locked to the orbital period, and
this in turn might imply planetary masses which are near the
minimum mass values (Drake et al. 1998).  

We expect 51~Peg~b and $\upsilon$~And~b to have larger radii
than the more massive $\tau$~Bo\"o~b (Guillot et al. 1996).  
If $\tau$~Bo\"o~b has been detected, it has a very large surface area.  
If 51~Peg~b and $\upsilon$~And~b have comparable albedos to $\tau$~Bo\"o~b,
they should have a comparable or larger reflected light ratio.

\section{The Transiting System HD~209458}
\begin{figure}
\plotone{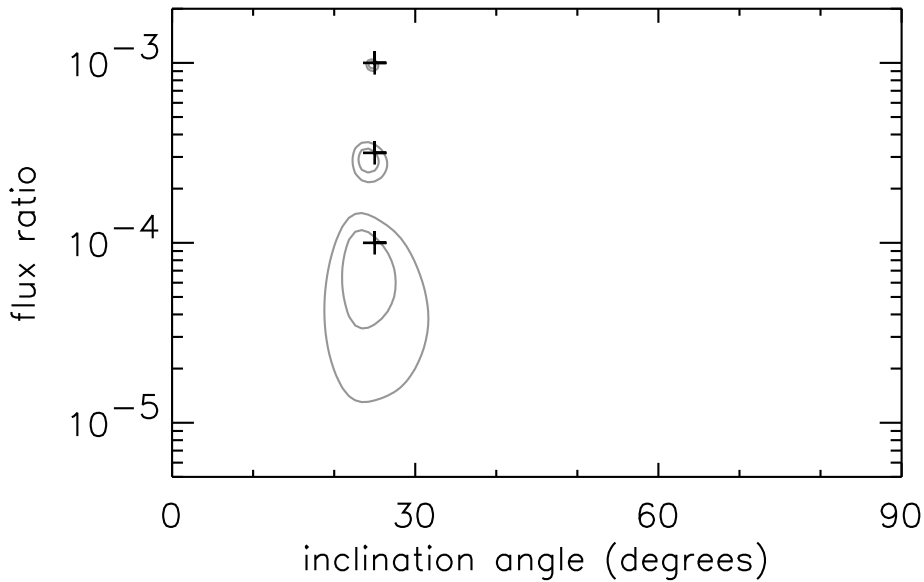}
\caption{The detection thresholds shown in Figure~1 
were tested by directly injecting a secondary into
the data at a given amplitude and inclination. Three such tests 
are shown above, and the black crosses
indicate the input parameters. The contours demarcate the
68\% and 95\% confidence regions on the derived parameters for the 
planet.}
\end{figure}

In the case of a planet that is observed to transit its star, the
situation improves considerably.  Since transit observations yield
$R_{p}$ and $i$, then only remaining unknowns in equation~(1) are
$p_{\lambda}$ and $\phi_{\lambda}(\alpha)$.

The first transiting extrasolar planet, HD~209458~b, was detected by
Charbonneau et al. (2000) and Henry et al. (2000).  The dominant uncertainty
in deriving the planetary radius and orbital inclination is the estimation
of the stellar radius, and not the photometric precision.  A detailed
analysis of the stellar modeling, and the resulting derivation of the
planetary parameters and the systematic uncertainties is presented
in Mazeh et al. (2000).  
They derive \mbox{$R_{p} = 1.40 \pm 0.17 \ R_{\rm Jup}$}, \mbox{$a=0.467$ AU}, 
and \mbox{$i=86\fdg1 \pm 1\fdg6$}.  
Substituting these values into equation~(3), we find 
$f_{\lambda}(\alpha) = 2.1 \times 10^{-4} \ p_{\lambda} \ \phi_{\lambda}(\alpha)$.

Transiting systems such as HD~209458 are particularly desirable
targets for reflected light measurements since by securing measurements
near opposition (where \mbox{$\phi_{\lambda}(0) \simeq 1$}),
an observed reflected light amplitude 
is a direct measurement of the geometric albedo.  
Furthermore, one can use the spectra obtained while the
planet is behind the star (rather than near inferior conjunction)
to create the required stellar template spectrum.  
One difficulty is that HD~209458 is significantly 
fainter \mbox{($V = 7.6$)} than the systems described above.

We have simulated what could be achieved in a modest observing run
using the Keck-1 telescope and HIRES spectrograph (Vogt et al. 1994),
based on experience gained from our earlier
observing run on $\tau$~Bo\"o (C99).  
In three nights when the planet is near opposition, 
we would obtain a detection threshold of \mbox{$4 \times 10^{-5}$} 
for the reflected flux ratio.  Thus we would constrain the 
geometric albedo to be \mbox{$p_{\lambda} \le 0.2$},
or directly detect the planet in reflected light.

\acknowledgements
We thank our collaborators in our reflected light study
of $\tau$~Bo\"o, Sylvain Korzennik, Peter Nisenson, Saurabh Jha,
Steven Vogt, and Robert Kibrick.  We also thank Timothy Brown for 
many discussions regarding HD~209458.

\end{document}